\documentclass[aip,jcp,reprint,showkeys]{revtex4-1}
\usepackage{bm,graphicx,tabularx,array,dcolumn,xcolor,microtype,multirow,amscd,amsmath,amssymb,amsfonts,physics,siunitx}
\usepackage[version=4]{mhchem}
\usepackage[utf8]{inputenc}
\usepackage[T1]{fontenc}
\usepackage{txfonts}
\usepackage{simpler-wick}

\usepackage[normalem]{ulem}
\definecolor{hughgreen}{RGB}{0, 128, 0}

\usepackage[
	colorlinks=true,
    citecolor=blue,
    linkcolor=blue,
    filecolor=blue,      
    urlcolor=blue,	
    breaklinks=true
	]{hyperref}
\usepackage{caption}
\usepackage{subcaption}

\usepackage[]{natbib}

\newcommand{\R}{{\mathbb{R}}}

\newcommand{\di}{{\rm d}}
\newcommand{\ee}{{\rm e}}

\newcommand{\rv}{{\bf r}}

\newcommand{\sca}{\gamma}

\newcommand{\bmath}{\begin{eqnarray}}
\newcommand{\emath}{\end{eqnarray}}

\newcommand{\AIMMS}{Department of Chemistry and Pharmaceutical Sciences, Amsterdam Institute of Molecular and Life Sciences (AIMMS), Faculty of Science, Vrije Universiteit, De Boelelaan 1083, 1081HV Amsterdam, The Netherlands}

\begin{document}

\title{Large-$Z$ atoms in the strong-interaction limit of DFT: Implications for gradient expansions and for the Lieb-Oxford bound}

\author{Kimberly J. \surname{Daas}}
\affiliation{\AIMMS}
\author{Derk P. \surname{Kooi}}
\affiliation{\AIMMS}
\affiliation{Microsoft Research AI4Science}
\author{Tarik~\surname{Benyahia}}
\affiliation{\AIMMS}
\author{Michael \surname{Seidl}}
\affiliation{\AIMMS}
\author{Paola \surname{Gori-Giorgi}}
\affiliation{\AIMMS}
\affiliation{Microsoft Research AI4Science}

\begin{abstract}
We study numerically the strong-interaction limit of the exchange-correlation functional for neutral atoms and for Bohr atoms as the number of electrons increases. Using a compact representation, we analyse the second-order gradient expansion, comparing it with the one for exchange (weak interaction limit).  The two gradient expansions, at strong and weak interaction, turn out to be very similar in magnitude, but with opposite signs. We find that the point-charge plus continuum model is surprisingly accurate for the gradient expansion coefficient at strong coupling, while generalized gradient approximations such as PBE and PBEsol severely underestimate it. We then use our results to analyse the Lieb-Oxford bound from the point of view of slowly-varying densities, clarifying some aspects on the bound at fixed number of electrons.
\end{abstract}

\maketitle
\section{Introduction}
Exact properties (or constraints) of the exchange-correlation (XC) functional of Kohn-Sham (KS) density functional theory (DFT) play a central role in the construction of practical approximations (see, e.g., refs.~\citenum{PerBurErn-PRL-96,PerRuzTaoStaScuCso-JCP-05,CohMorYan-CR-12,PerRuzSunBur-JCP-14,PerConSagBur-PRL-06,PerRuzCsoVydScuConZhoBur-PRL-08,SunRuzPer-PRL-15,MegaPaper-PCCP-22}). Many of them have been derived and incorporated into useful approximations by John Perdew and his coworkers. In particular, the slowly-varying limit,\cite{AntKle-PRB-85,KleLee-PRB-88,SvevBa-IJQC-95,vLe-PRB-13}  already invoked in the seminal paper of Kohn and Sham,\cite{KohSha-PR-65} and the subsequent
 generalised gradient approximations (GGA's) made KS DFT the workhorse for computational chemistry and solid-state physics.\cite{MegaPaper-PCCP-22} 
The crucial role played by John Perdew in this success cannot be overstated.\cite{PerBurErn-PRL-96,PerRuzCsoVydScuConZhoBur-PRL-08}

While successful GGA's for solids\cite{PerRuzCsoVydScuConZhoBur-PRL-08} typically recover the exact\cite{AntKle-PRB-85,KleLee-PRB-88,SvevBa-IJQC-95,vLe-PRB-13} second-order gradient expansion coefficient  for exchange, chemical systems are better described with GGA's with a coefficient almost twice as large in magnitude.\cite{Bec-PRA-88,PerBurErn-PRL-96} This empirical observation was later rationalised by connecting the gradient expansion with the large-$Z$ limit of neutral atoms, where $Z$ is the nuclear charge.\cite{PerConSagBur-PRL-06,EllBur-CJC-09} The argument partially relied on an assumption on the large-$Z$ dependence of exchange beyond the local density approximation (LDA), which was only very recently corrected.\cite{DaaKooGroSeiGor-JCTC-22,ArgRedCanBur-PRL-22}

The exchange functional is the weakly-interacting (or high-density) limit of the exact XC functional. The opposite limit, strongly interacting (SIL) or low density,\cite{SeiGorSav-PRA-07,GorVigSei-JCTC-09,VucGerDaaBahFriGor-WIREs-22} provides complementary information,  and can be used to build approximations in different ways (for a recent review see ref.~\citenum{VucGerDaaBahFriGor-WIREs-22}). The main aim of this work is to study the SIL functional for large-$Z$ atoms, computing accurate numerical results and providing an analysis similar to the one done for exchange.\cite{EllBur-CJC-09,ArgRedCanBur-PRL-22} To fully take into account the recent corrections on the large-$Z$ behavior,\cite{DaaKooGroSeiGor-JCTC-22,ArgRedCanBur-PRL-22} we analyse our results through the compact representation introduced in previous work on the strong-coupling limit of the M\o ller-Plesset adiabatic connection.\cite{DaaKooGroSeiGor-JCTC-22} We also perform this kind of compact analysis on exchange, which reveals a somewhat surprising symmetry between the two limits: the resulting gradient expansions are very similar in magnitude, but with opposite signs. As we shall see, our numerical study on neutral atoms and Bohr atoms also suggests a weak dependence on density profiles of this gradient expansion of both limits, in agreement with Ref.~\onlinecite{ArgRedCanBur-PRL-22}. We also compare our accurate results with different approximations.

We then turn to another important exact constraint for the XC functional: the Lieb-Oxford (LO) inequality,\cite{Lie-PLA-79,LieOxf-IJQC-81,LewLieSei-LMP-22,PerSun-arxiv-22} which has been turned into a useful tool for constraining approximations, again, by John Perdew.\cite{Per-INC-91,PerSun-arxiv-22} 

In previous works,\cite{RasSeiGor-PRB-11,SeiVucGor-MP-16,SeiBenKooGor-BOOK-22} the SIL functional has been used to establish lower bounds for the optimal constant appearing in the LO inequality at given electrons number $N$.\cite{RasSeiGor-PRB-11,SeiVucGor-MP-16,SeiBenKooGor-BOOK-22} A question that remained open in this context was why some density profiles give much tighter bounds than others. We will answer to this question by using the present results, and we will also investigate the relation with the functional appearing in the strong-coupling limit of the M\"oller-Plesset adiabatic connection.\cite{SeiGiaVucFabGor-JCP-18,DaaGroVucMusKooSeiGieGor-JCP-20,DaaKooGroSeiGor-JCTC-22}

Hartree atomic units are used throughout.

\section{Theoretical background}
\label{sec:theory}
For a given $N$-electron density $\rho(\rv)$, the Levy\cite{Lev-PNAS-79}-Lieb\cite{Lie-IJQC-83} universal functional for general interaction strength $\lambda$ is defined as
\begin{equation}
    F_\lambda[\rho]=\min_{\Psi\mapsto \rho}\langle\Psi|\hat{T}+\lambda\,\hat{V}_{ee}|\Psi\rangle,
\end{equation}
where $\hat{T}$ is the kinetic energy operator for the $N$ electrons, $\hat{V}_{ee}$ is their mutual Coulomb repulsion, and the minimization is performed over all many-electron wavefunctions with the prescribed density $\rho(\rv)$. The XC functional that needs to be approximated in any practical KS DFT calculation is
\begin{equation}
    E_{\rm xc}[\rho]=F_1[\rho]-F_0[\rho]-U[\rho],
\end{equation}
where $U[\rho]$ is the Hartree (mean field) functional,
\begin{equation}
    U[\rho]=\frac{1}{2}\int d\rv\int d\rv' \frac{\rho(\rv)\rho(\rv')}{|\rv-\rv'|}.
\end{equation}
\subsection{The functionals $E_{\rm x}[\rho]$ and $W_\infty[\rho]$}
For the exact XC functional, applying uniform coordinate scaling by defining $\rho_\sca(\rv)=\sca^3\rho(\sca\rv)$,
\bmath
\int\di\rv\,\rho_\sca(\rv)\;=\;N\qquad\text{(for all $\sca>0$)},
\emath
is equivalent\cite{LevPer-PRA-85,LevPer-PRB-93} to scale the strength of the electron-electron interaction, $\lambda=1/\sca$.
The high ($\lambda\to0$) and low ($\lambda\to\infty$) density limits (weakly and strongly interacting limits, respectively) of the functional $E_{\rm xc}[\rho]$ are known to be 
\bmath
\lim_{\lambda\to0}\Big(\lambda\,E_{\rm xc}[\rho_{1/\lambda}]\Big) &=& W_{0}[\rho]  \quad\equiv\quad E_{\rm x}[\rho],  \nonumber \\
 \lim_{\lambda\to\infty}\Big(\lambda\,E_{\rm xc}[\rho_{1/\lambda}]\Big) &=& W_{\infty}[\rho]. 
\label{WiExDef}
\emath
The high density ($\frac1{\lambda}\to\infty$) limit $W_{0}[\rho]=E_{\rm x}[\rho]$ is also called the DFT exchange energy. Notice that both functionals satisfy
\begin{equation}\label{eq:scarel}
    E_{\rm x}[\rho_\sca]=\sca E_{\rm x}[\rho],\qquad W_{\infty}[\rho_\sca]=\sca W_{\infty}[\rho].
\end{equation}

If we look at the spin-polarization dependence, considering the spin-densities $\rho_\uparrow(\rv)$ and $\rho_\downarrow(\rv)$, with $\rho=\rho_\uparrow+\rho_\downarrow$, we have~\cite{OliPer-PRA-79,SeiPerKur-PRA-00,KapLevPer-Arx-22}
\begin{align} \label{eq:spinEx}
    E_{\rm x}\left[\rho_\uparrow,\rho_\downarrow\right] & =\frac{1}{2}E_{\rm x}\left[2\rho_\uparrow\right]+\frac{1}{2}E_{\rm x}\left[2\rho_\downarrow\right], \\
W_\infty\left[\rho_\uparrow,\rho_\downarrow\right] & =W_\infty[\rho]. \label{eq:spinWinf}
\end{align}
The spin-independence of the functional $W_\infty[\rho]$ is due to the fact that, as $\lambda\to\infty$, electrons are strictly correlated, forming a floating crystal in a metric dictated by the density $\rho(\rv)$, with spin effects appearing at orders $\sim e^{-\sqrt{\lambda}}$.\cite{SeiGorSav-PRA-07,GorVigSei-JCTC-09,GroKooGieSeiCohMorGor-JCTC-17}

In the rest of this paper, we will consider closed-shell systems only, with $\rho_\uparrow=\rho_\downarrow=\rho/2$. {\color{black}Our results for both functionals can be extended to cases with $\rho_\uparrow \neq \rho_\downarrow$ via Eqs.\eqref{eq:spinEx}-\eqref{eq:spinWinf}.}

\subsection{Gradient expansion of $E_{\rm x}[\rho]$ and $W_\infty[\rho]$}
Central to many approximate XC functionals is the slowly varying limit, in terms of gradients of the density. The scaling relations of Eq.~\eqref{eq:scarel} imply that if a gradient expansion approximation (GEA) for the two functionals $W_i[\rho]$ ($i=0$ or $\infty$) of Eq.~\eqref{WiExDef}  exists, its first two leading terms 
must be expressed using the integrals
\bmath
I_0[\rho] &=& \int\di\rv\,\rho(\rv)^{4/3}, \label{eq:I0}\\
I_2[\rho] &=& \int\di\rv\,\frac{\,|\nabla\rho(\rv)|^2}{\rho(\rv)^{4/3}}, \label{eq:I2}
\emath
and must have the form
\bmath
W_i[\rho] &=& \underbrace{\overbrace{A_i\cdot I_0[\rho]}^{W_i^{\rm LDA}[\rho]}
\;+\;B_i\cdot I_2[\rho]}_{W_i^{\rm GEA2}[\rho]}\,+\,...,\nonumber\\
&=&A_i\int\di\rv\,\rho(\rv)^{4/3}\left(1\,+\,\frac{B_i}{A_i} \cdot x\big([\rho],\rv\big)^2\right)\,+\,....
\label{generalGEA}\emath
In the second line, we have introduced the reduced gradient
\bmath
x\big([\rho],\rv\big)\;=\;\frac{\,|\nabla\rho(\rv)|}{\rho(\rv)^{4/3}},
\emath
which essentially gives the relative change of the density on the scale of the average interparticle distance. In the DFT literature the equivalent reduced gradient  $s([\rho],\rv)=\frac{1}{2}(3\pi^2)^{-1/3}x([\rho],\rv)$ is often used, as it describes more accurately the relevant length scale for exchange when perturbing an infinite system of uniform density. 

{\color{black} For the case of exchange ($i=0$), in Ref.~\citenum{SprSveBar-PRB-96} it was shown that if a gradient expansion exists, it is an asymptotic series: earlier cut-offs in the series are needed as the strength of the potential perturbing the constant density increases. For the strong interaction limit ($i=\infty$), even less is known on the existence of a gradient expansion. However, on a purely practical side, GEA~\cite{SeiPerKur-PRA-00} and GGA~\cite{SmiDelGorFab-JCTC-22} functionals have been shown, by using available SIL results on atoms, to accurately approximate the first two leading order terms of the strong interaction limit. These gradient expansions have even been used in interpolations functionals that accurately describe a variety of larger systems.~\cite{FabGorSeiDel-JCTC-16,GiaGorDelFab-JCP-18,VucGorDelFab-JPCL-18,SmiDelGorFab-JCTC-22}}

The LDA constants $A_i$ are exactly known, while the value of the coefficients $B_i$ is more subtle. By applying a slowly varying perturbation to the uniform electron gas, and by carefully handling the long-range Coulomb tail, a coefficient $B_{\rm x}^{\rm GEA}$ for the exchange functional has been derived\cite{AntKle-PRB-85,KleLee-PRB-88,SvevBa-IJQC-95,vLe-PRB-13} (we use equivalently $i=0$ or $i={\rm x}$ to denote quantities for the functional $W_0=E_{\rm x}$). However, while GGA's that recover $B_{\rm x}^{\rm GEA}$ work well for extended systems, for atoms and molecules better results are obtained with a value roughly twice as large in magnitude. In Table~\ref{tab:values} we report a small overview, with some values of $B_{\rm x}$ for GGA exchange functionals widely used in chemistry.

For $W_\infty[\rho]$, the value of $B_\infty$ is unknown. The point-charge plus continuum (PC) model,\cite{SeiPerKur-PRA-00} provides an approximate value for this coefficient, as well as an approximate value for $A_\infty$, which is slightly different than the exact one. The PC model is constructed from the physical idea that for slowly varying densities the electrons will try to neutralise the small dipole created by the density gradient. The PC values are also reported in Table~\ref{tab:values}, together with the $B_\infty$ from the Perdew-Burke-Ernzerhof (PBE)\cite{PerBurErn-PRL-96} and the PBEsol\cite{PerRuzCsoVydScuConZhoBur-PRL-08} XC functionals. 

Gaining more information on the coefficient $B_\infty$ is one of the aims of this article. To this purpose, in the following section \ref{sec:partnum} we will review particle number scaling\cite{PerConSagBur-PRL-06} and the large-$Z$ limit of atoms\cite{PerConSagBur-PRL-06,ArgRedCanBur-PRL-22}  as an alternative way to approach the slowly-varying limit for finite systems. Notice that this procedure is different than perturbing the uniform electron gas. While both procedures are expected to yield the same coefficients $A_i$,\cite{LewLieSei-PAA-20} studies on the exchange functional\cite{PerConSagBur-PRL-06,ArgRedCanBur-PRL-22} found that the coefficient $B_i$ is not the same, which could explain\cite{PerConSagBur-PRL-06,EllBur-CJC-09,ArgRedCanBur-PRL-22}  why successful GGA's for chemistry have $B_{\rm x}\neq B_{\rm x}^{\rm GEA}$, as exemplified in Table~\ref{tab:values}, where we also report the value $B_{\rm x}^{\rm ARCB}$ recently extracted from the large-$Z$ limit of atoms.\cite{ArgRedCanBur-PRL-22}

\begin{table*}
\begin{tabular}{|l||l|l|}\hline
$W_i[\rho]$ & $A_i$ & $B_i$ \\\hline\hline
$E_{\rm x}[\rho]$ & $A_{\rm x}=-\frac34\big(\frac3{\pi}\big)^{1/3}\approx -0.73856$
                          & $B^{\rm GEA}_{\rm x}=-\frac{5}{216\,\pi\,(3\pi^2)^{1/3}}\approx -0.0024$ 	\\
                    &   &$B_{\rm x}^{\rm PBE}=-0.0042$ \\
                  &       & $B_{\rm x}^{\rm B88}=-0.0053$	\\
                &       & $B_{\rm x}^{\rm ARCB}=-\frac{1}{16\, \pi\,(3 \pi^2)^{1/3}}\approx-0.0064$
	\\\hline
 $W_\infty[\rho]$  & $A_\infty=-1.44423075$ & $B_\infty^{\rm GEA}=?$ \\
                  & $A_\infty^{\rm PC}=-\frac9{10}\big(\frac{4\pi}3\big)^{1/3}\approx-1.4508$
				  & $B_{\infty}^{\rm PC}=\frac3{350}\big(\frac3{4\pi}\big)^{1/3}\approx 0.0053173$	\\
				&       & $B_{\infty}^{\rm PBE}=1.24457\cdot 10^{-7}$\\
                &       & $B_{\infty}^{\rm PBEsol}=0.0005378$
	\\\hline
\end{tabular}
\caption{Coefficients $A_i$ and $B_i$ appearing in Eq.~\eqref{generalGEA}. The value of $A_\infty$ is given by the bcc Wigner crystal energy, which is floating to recover the uniform density.\cite{LewLieSei-PRB-19} Values for the point-charge plus continuum (PC) model\cite{SeiPerKur-PRA-00} and for the PBE,\cite{PerBurErn-PRL-96} PBEsol\cite{PerRuzCsoVydScuConZhoBur-PRL-08} and B88\cite{Bec-PRA-88} functionals are also shown. The value $B^{\rm ARCB}_{\rm x}$ has been extracted very recently from an accurate study of large-$Z$ neutral and Bohr atoms.\cite{ArgRedCanBur-PRL-22}}
\label{tab:values}
\end{table*}

\section{Gradient expansion from particle-number scaling}
\label{sec:partnum}
The gradient expansion of Eq.~\eqref{generalGEA} is usually invoked for densities $\rho$ with weak reduced gradient,
\bmath
x\big([\rho],\rv\big)\;\ll\;1.
\label{redGradCond}
\emath
While perturbing a uniform density is one way to create such densities, adding more and more particles in a fixed density profile is another possibility, as detailed below. 

\subsection{Particle-number scaling of a density profile}
For a given density profile $\bar{\rho}(\rv)$, with $\int\di\rv\,\bar{\rho}(\rv)=1$, (and for a given exponent $p$)
we construct (``particle-number scaling'') a sequence of densities,
\bmath
\bar{\rho}_{N,p}(\rv)\;=\;N^{3p+1}\,\bar{\rho}(N^p\,\rv),
\label{rhoPNsca}\emath
with increasing particle number $N$,
\bmath
\int\di^3r\,\bar{\rho}_{N,p}(\rv)\;=\;N\qquad(N=1,2,3,...).
\emath
These densities have the reduced gradient
\bmath
x\big([\bar{\rho}_{N,p}],\rv\big)\;=\;\frac{x\big([\bar{\rho}],N^p\rv\big)}{N^{1/3}}.
\emath
Consequently, for sufficiently large particle numbers $N\gg1$, they satisfy condition \eqref{redGradCond}, provided that
\bmath
\label{eq:finitegra}
\max_{\rv\in\R^3} x\big([\bar{\rho}],\rv\big)\qquad\text{is finite.}
\emath
Exponentially decaying densities (or other kinds of $L^2$ densities) do not satisfy Eq.~\eqref{eq:finitegra} in their tails. In these cases, one applies the GEA in the sense of the right-hand-side of the first line of Eq.~\eqref{generalGEA}, with the second term being much smaller than the first one. 

Using the densities \eqref{rhoPNsca} in Eq.~\eqref{generalGEA}, we see that the existence of a gradient expansion for the functionals $E_{\rm x}[\rho]$ and $W_\infty[\rho]$ implies a well defined  large-$N$ expansion  
\bmath
W_i\big[\bar{\rho}_{N,p}\big] &=& A_i\cdot I_0\big[\bar{\rho}_{N,p}\big]\;+\;B_i\cdot I_2\big[\bar{\rho}_{N,p}\big]\;+\;...\nonumber\\
&=& A_i\cdot I_0[\bar{\rho}]\cdot N^{p+4/3}\;+\;B_i\cdot I_2[\bar{\rho}]\cdot N^{p+2/3}\;+\;....
\label{WexplicitN}\emath
Provided that the terms indicated by dots are sufficiently small, we may conclude
\bmath
\frac{W_i\big[\bar{\rho}_{N,p}\big]\;-\;A_i\cdot I_0[\bar{\rho}]\cdot N^{p+4/3}}{I_2[\bar{\rho}]\cdot N^{p+2/3}} &=& B_i\,+\,O(N^{-1/3}), \nonumber \\
\lim_{N\to\infty}\frac{W_i\big[\bar{\rho}_{N,p}\big]\;-\;A_i\cdot I_0[\bar{\rho}]\cdot N^{p+4/3}}{I_2[\bar{\rho}]\cdot N^{p+2/3}} &=& B_i. \label{eq:limitBi}
\emath
Notice that the uniform-coordinate scaling of Eq.~\eqref{eq:scarel} implies that all choices of $p$ are  equivalent for studying $E_{\rm x}[\rho]$ and $W_\infty[\rho]$, as it holds
\begin{equation}
\label{eq:pdep}
W_i\big[\bar{\rho}_{N,p}\big]=N^p W_i\big[\bar{\rho}_{N,0}\big].
\end{equation}
For functionals that do not satisfy a simple relation under uniform coordinate scaling, such as the correlation functional, different values of $p$ explore different physical regimes, as reviewed in Ref.~\citenum{FabCon-PRA-13}.

\subsection{Densities with asymptotic particle-number scaling}
Physical many-electron systems  do not arise by filling more and more particles in a fixed density profile, but by adding particles in an external potential. In order to study the gradient expansion for physically relevant systems, several authors considered neutral atoms,\cite{Lie-RMP-81,PerConSagBur-PRL-06,EllBur-CJC-09,ArgRedCanBur-PRL-22} in which $N=Z$ electrons are bound by a point charge $Z$, and so-called Bohr atoms,\cite{HeiLie-PRA-95,KapSanBhaWagChoBheYuTanBurLevPer-JCP-20,ArgRedCanBur-PRL-22} in which the external potential is $-1/r$ and the electron-electron interaction is set to zero. These systems define
a sequence of $N$-electron densities $\rho^{\rm Sqc}_{N}(\rv)$,
\bmath
\int\di\rv\,\rho^{\rm Sqc}_{N}(\rv)\;=\;N\qquad(N=1,2,3,...),
\emath
which displays particle-scaling behavior only asymptotically (in the limit of large $N\gg1$),
\bmath
\rho^{\rm Sqc}_{N}(\rv) &\approx& \bar{\rho}^{\rm Sqc}_{N,p}(\rv)\qquad\qquad(N\gg1)\\
&=& N^{3p+1}\,\bar{\rho}^{\rm Sqc}(N^p\,\rv),
\label{rhoPNasy}
\emath
where $\bar{\rho}^{\rm Sqc}(\rv)$ is an asymptotic density profile, specific for the sequence $\rho^{\rm Sqc}_{N}(\rv)$, which can be obtained exactly from Thomas-Fermi (TF) theory.\cite{Lie-RMP-81,HeiLie-PRA-95,OkuBur-BOOK-23,KapSanBhaWagChoBheYuTanBurLevPer-JCP-20,ArgRedCanBur-PRL-22}

In this case, using the densities $\rho^{\rm Sqc}_{N}(\rv)$ in Eq.~\eqref{generalGEA}, we obtain, for large $N$,
\bmath
W_i\big[\rho^{\rm Sqc}_{N}\big]
\;=\; A_i\cdot I_0\big[\rho^{\rm Sqc}_{N}\big]\;+\;B_i\cdot I_2\big[\rho^{\rm Sqc}_{N}\big]+\dots
\label{GEAasy}\emath
Unlike in Eq.~\eqref{WexplicitN}, however, the $N$-dependence cannot be extracted explicitly here. Moreover, even when the integrals
$I_{0,2}\big[\rho^{\rm Sqc}_{N}\big]$ are finite for all values of $N$, the corresponding integrals
$I_{0,2}\big[\bar{\rho}^{\rm Sqc}\big]$ for the asymptotic profile $\bar{\rho}^{\rm Sqc}(\rv)$ can be divergent,
as we shall see below.

\subsubsection{Neutral Atoms}
\label{sub:na}
We consider here the densities of neutral atoms (writing "na" for "Sqc"), $\rho^{\rm Sqc}_{N}(\rv)\;=\;\rho^{\rm na}_{N}(\rv)$.
In this case, we have asymptotically, as $N$ gets larger and larger,\cite{Lie-RMP-81,LeeConPerBur-JCP-09,OkuBur-BOOK-23} 
\bmath
\rho^{\rm na}_{N}(\rv) &\approx& \bar{\rho}^{\rm TFna}_{N,1/3}(\rv)\nonumber\\
&=& N^{2}\,\bar{\rho}^{\rm TFna}(N^{1/3}\,\rv)\qquad(p=\textstyle\frac13).
\label{scaHFna}
\emath
The Thomas-Fermi profile $\bar{\rho}^{\rm TFna}(\rv)$ does not have a closed form, but a very accurate parametrization is provided in Ref.~\citenum{LeeConPerBur-JCP-09}. While $I_0[\bar{\rho}^{\rm TFna}]$ has a finite value, $I_2[\bar{\rho}^{\rm TFna}]$ diverges.
This latter divergence has consequences for the large $N$ (or, equivalently, large-$Z$) behavior of $E_{\rm x}[\rho^{\rm na}_{N}]$, which was somehow overlooked in earlier works\cite{PerConSagBur-PRL-06,EllBur-CJC-09,PerRuzCsoVydScuConZhoBur-PRL-08}, and was recently reconsidered.\cite{DaaKooGroSeiGor-JCTC-22,ArgRedCanBur-PRL-22} In particular, we have the large-$N$ asymptotics\cite{DaaKooGroSeiGor-JCTC-22,ArgRedCanBur-PRL-22}
\bmath
I_0\big[\rho^{\rm na}_{N}\big] &=& a_0^{\rm na} N^{5/3}+a_1^{\rm na}\,N\log(N)+\dots \label{eq:I0na} \\
I_2\big[\rho^{\rm na}_{N}\big] &=& b_1^{\rm na}\,N\log (N)\;+b_2^{\rm na}\,N+\dots \label{eq:I2na} 
\emath
 
\subsubsection{Bohr atoms}
The (closed shell) Bohr atoms densities (writing "Bohr" for "Sqc") are given by
\bmath
\rho^{\rm Bohr}_{N}(\rv)\;=\;2\sum_{n=1}^{k_N}\sum_{\ell=0}^{n-1}\sum_{m_\ell=-\ell}^{\ell}\big|\psi_{n\ell m_\ell}(\rv)\big|^2,
\emath
with the hydrogenic orbitals $\psi_{n\ell m_\ell}(\rv)=R_{n\ell}(r)\,Y_{\ell m_\ell}(\theta,\phi)$ and $k_2=1$, $k_{10}=2$, $k_{28}=3$, ...

Atomic ions with $N$ {\bf non-interacting} electrons (NIE) and nuclear charge $Z$ obviously have in their ground state exactly the electron density
\bmath \label{eq:NIEions}
\rho^{\rm IonNIE}_{Z,N}(\rv)\;=\;Z^3\,\rho^{\rm Bohr}_{N}(Z\,\rv)\qquad\qquad\text{(NIE)}.
\emath
The exact density $\rho^{\rm Ion}_{Z,N}(\rv)$ of an {\bf interacting} (non relativistic) atomic ion with $Z\gg N$ asymptotically approaches the NIE one,
\bmath
\rho^{\rm Ion}_{Z,N}(\rv)\;\to\;\rho^{\rm IonNIE}_{Z,N}(\rv)\qquad\qquad(Z\gg N).
\emath

The Bohr atom densities satisfy asymptotic particle-number scaling with $p=-2/3$,
\bmath
\rho^{\rm Bohr}_{N}(\rv) &\approx& \bar{\rho}^{\rm TFBohr}_{N,\,-2/3}(\rv) \;=\; \frac1{N}\,\bar{\rho}^{\rm TFBohr}(N^{-2/3}\,\rv).
\label{scaBohr}
\emath

The TF profile $\bar{\rho}^{\rm TFBohr}(\rv)$ has a simple closed form that is reported, for example, in Refs.~\citenum{OkuBur-BOOK-23,KapSanBhaWagChoBheYuTanBurLevPer-JCP-20}. As for neutral atoms, $I_0[\bar{\rho}^{\rm TFBohr}]$ is finite while $I_2[\bar{\rho}^{\rm TFBohr}]$ diverges. The divergence of $I_2$ for Bohr atoms has been carefully analysed by Argaman et al.\cite{ArgRedCanBur-PRL-22}

\section{Compact representation to study the gradient expansion}
\label{sec:gradexp}
Previous works that used large-$Z$ (or large-$N$) neutral and Bohr atoms data ("na" or "Bohr" for "Sqc" in our notation) to extract the coefficient $B_{\rm x}$ numerically, fitted the $N$-dependence of the exchange energy $W_0[\rho]=E_{\rm x}[\rho]$ (or the difference
between $E_{\rm x}[\rho]$ and its LDA counterpart, $A_{\rm x}\,I_0[\rho]$) for large $N$.\cite{EllBur-CJC-09,ArgRedCanBur-PRL-22} This procedure relies on knowledge  of the large-$N$ behaviour (see Appendix~\ref{app:Bohr}), which, in view of the diverging nature of $I_2[\rho_N^{\rm Sqc}]$, can easily lead to erroneous assumptions.\cite{PerRuzCsoVydScuConZhoBur-PRL-08,EllBur-CJC-09,DaaKooGroSeiGor-JCTC-22,ArgRedCanBur-PRL-22}  
Moreover, a separate fit for each sequence (Sqc) needs to be done.

Here we rely on a different procedure:\cite{DaaKooGroSeiGor-JCTC-22}  Using numerical densities $\rho(\rv) = \rho_N^{\rm Sqc}(\rv)$ and energies $W_i[\rho_N^{\rm Sqc}]$ (where $i={\rm x}$ or $i=\infty$), we compute for various sequences (Sqc) and increasing particle numbers $N$ the values
\begin{equation}
\label{eq:BiN}
\frac{W_i[\rho_N^{\rm Sqc}]-A_i\cdot I_0[\rho_N^{\rm Sqc}]}{I_2[\rho_N^{\rm Sqc}]} \;=\; \tilde{B}^{\rm Sqc}_i(N).
\end{equation}
For $N\to\infty$, each sequence $\rho_N^{\rm Sqc}(\rv)$ approaches the limit of a slowly varying density. Therefore, if the
GEAs are valid for the functionals $W_i[\rho]$, the numbers $\tilde{B}^{\rm Sqc}_i(N)$ will approach the sought coefficients $B_i$, as explicitly shown in Eq.~\eqref{eq:limitBi} for a scaled profile (notice that the constant $B_
i$ is expected to be approached  slower than $\propto N^{-1/3}$ for neutral and Bohr atoms). Whether the $B_i$ will be the same for all sequences (i.e., whether they are profile-independent, and whether they are the same for a scaled profile and for neutral and Bohr atoms) is an open question, but this approach allows us to use data from different sequences to address this point more easily than the approach based on fitting the $N$-dependence of the energy. It also allows us to combine data obtained from scaled density profiles and data obtained from neutral and Bohr atoms, as the leading-order $N$-dependence of numerator and denominator (whether linear or with logarithmic terms in $N$) will cancel if the GEA's are valid.

\begin{figure}
\includegraphics[width=\columnwidth]{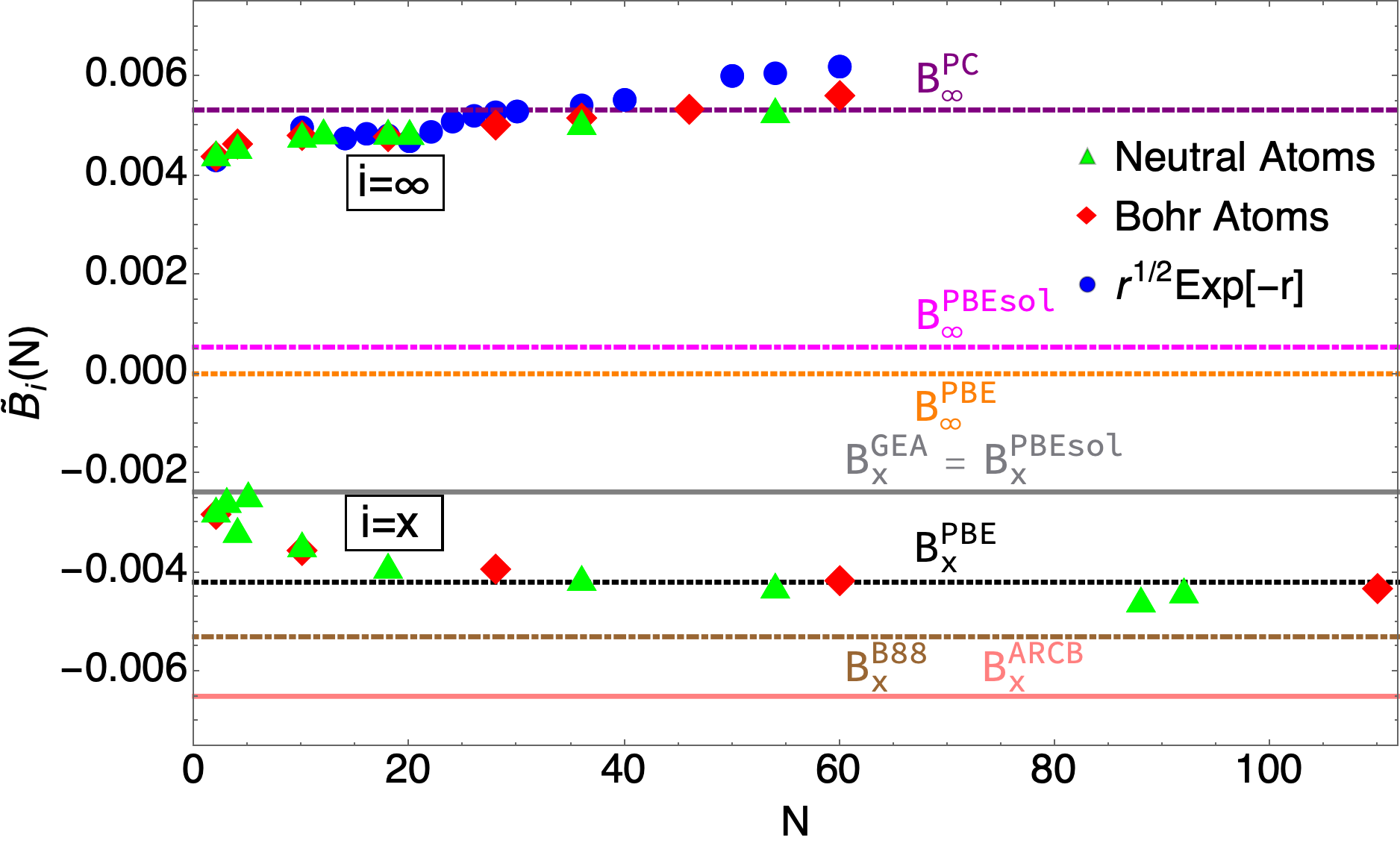}
\caption{Numerical values for $\tilde{B}^{\rm Sqc}_{\rm x}(N)$ and $\tilde{B}^{\rm Sqc}_\infty(N)$ of Eq.~\eqref{eq:BiN} for different sequences of densities:  {\color{green} Green: Neutral atoms.} {\color{red} Red: Bohr atoms.}
{\color{blue} Blue: $\rho^{\rm Sqc}_N(\rv)=\frac{2N}{15\,\pi^{3/2}}\,\sqrt{r}\,\ee^{-r}$.}
The values for $B_i$ of Table~\ref{tab:values} from the literature are also shown. {\color{black}HF densities were used for the Neutral atoms.}}
\label{fig:Bi}
\end{figure}

\subsection{Densities}
The systems we have considered to generate data for $\tilde{B}_i^{\rm Sqc}(N)$ are:
\begin{itemize}
\item Closed-shell neutral atoms, treated at the Hartree-Fock level;
\item Closed-shell Bohr atoms;
\item  Only for $B_\infty^{\rm Sqc}(N)$: the particle-number scaled profile $\propto \sqrt{r}\,e^{-r}$, which was also used in Ref.~\citenum{SeiVucGor-MP-16}.
\end{itemize}
The full computational details are reported in Sec.~\ref{sec:compdet}. {\color{black} Although HF densities were used for the neutral atoms, we do not expect much difference between these results and results coming from Optimized-Effective-Potential (OEP) densities, see Sec~\ref{sec:Exfunc}.} We should also immediately mention that generating accurate data for $W_\infty[\rho]$ with large particle numbers is very challenging. For this reason, data for $B_\infty^{\rm Sqc}(N)$ are limited to $N\leq 60$.
\subsection{Results}
Our results for $B_i^{\rm Sqc}(N)$ are reported in Fig.~\ref{fig:Bi}, together with the various values for $B_i$ from literature of Table~\ref{tab:values}. 

The figure shows a surprising symmetry: the two extreme limits of correlation for the XC functional seem to have very similar effective gradient expansions in magnitude, but with opposite signs. 

The second interesting feature is that the profile-dependence seems rather small, giving some hope for the existence of a universal gradient expansion for finite systems. Regarding exchange, we do not have data with a fixed scaled profile (which would require KS inversion techniques), and thus we do not know whether neutral and Bohr atoms give similar results because of their similar asymptotic diverging behavior of the GEA integral.\cite{ArgRedCanBur-PRL-22} 

We should stress that our data are limited to relatively small numbers of electrons, and that the asymptotic value of $\tilde{B}_i^{\rm Sqc}(N)$ is approached very slowly (as $N^{-1/3}$ for scaled profiles, and even slower for neutral and Bohr atoms), which means that although the data look reasonably flat, the asymptotic value is probably still further out. Indeed, this seems to be confirmed by the value $B_{\rm x}^{\rm ARCB}$, also shown in the figure, which was extracted in Ref.~\citenum{ArgRedCanBur-PRL-22} from data for the Bohr atoms at much larger $N$. However, we also see in the figure that very successful GGA's as PBE and B88 have $B_{\rm x}$ values close to our data, suggesting that the chemically relevant region is in the range we are considering here rather than the final $N\to\infty$ limit.  This is further discussed in Appendix~\ref{app:Bohr}.

For $B_\infty$, we see that the PC model\cite{SeiPerKur-PRA-00} is surprisingly good, especially considering its fully non-empirical derivation, which was based on strictly-correlated electrons in an almost uniform density. The fact that it works so well for finite systems is certainly remarkable. The values obtained from the PBE and PBEsol functionals are way too small.
Notice that our numerical values $\tilde{B}^{\rm Sqc}_{\infty}(N)>0$ are variational and therefore possibly slightly too high.\cite{SeiDiMGerNenGieGor-arxiv-17}

For $B_{\rm x}$, our Fig.~\ref{fig:Bi} confirms previous studies on gradient expansion and the large-$Z$ neutral and Bohr atoms,\cite{PerConSagBur-PRL-06,EllBur-CJC-09,ArgRedCanBur-PRL-22}  which can now be easily visualised together using our $\tilde{B}_i^{\rm Sqc}(N)$. 

\section{Implications for the Lieb-Oxford bound}
The LO inequality\cite{Lie-PLA-79,LieOxf-IJQC-81,LewLieSei-LMP-22} provides a lower bound for the XC energy in terms of the integral $I_0[\rho]$ of Eq.~\eqref{eq:I0}, and has been used as exact constraint in many successful approximations for the XC functional.\cite{PerBurErn-PRL-96,SunRuzPer-PRL-15,PerSun-arxiv-22}. Including the two functionals $E_{\rm x}[\rho]$ and $W_\infty[\rho]$, the LO bound implies a chain of inequalities,
\bmath\label{eq:liebo}
 -C\int\di^3r\,\rho(\rv)^{4/3} \le W_\infty[\rho] \le E_{\rm xc}[\rho] \le E_{\rm x}[\rho] \le 0 ,
\emath
where the optimal value of the positive constant $C$ satisfies\cite{LewLieSei-LMP-22}
\begin{equation}\label{eq:Cbounds}
    \underbrace{1.44423}_{|A_\infty|} \le C \le 1.5765.
\end{equation}
Dividing Eq.~\eqref{eq:liebo} by $-I_0[\rho]$, we obtain
\bmath
\label{eq:LambdaLO}
0\le\frac{-E_{\rm x}[\rho]}{I_0[\rho]}\le\frac{-E_{\rm xc}[\rho]}{I_0[\rho]}\le\underbrace{\frac{-W_{\infty}[\rho]}{I_0[\rho]}}_{\Lambda_C[\rho]}\le C.
\emath
(Equivalently, the functional $\Lambda[\rho]=\frac1{|A_{\rm x}|}\Lambda_C[\rho]$ has been also used to analyse the LO bound in previous works\cite{RasPitCapPro-PRL-09,RasSeiGor-PRB-11,SeiVucGor-MP-16}).

The lower bound $|A_\infty|$ for $C$ in Eq.~\eqref{eq:Cbounds} is the highest value of the functional $\Lambda_C[\rho]$ in Eq.~\eqref{eq:LambdaLO} ever observed: a floating bcc Wigner crystal with uniform one-electron density.\cite{LewLieSei-PRB-19} The upper bound has been proven in Ref.~\citenum{LewLieSei-LMP-22}.

Lieb and Oxford\cite{LieOxf-IJQC-81} have also proven that if in Eq.~\eqref{eq:liebo} we consider only densities with a fixed number of electrons $N$, there is an optimal constant $c(N)$ for each $N$, and that $c(N)\le c(N+1)$. 

The functional $\Lambda_C[\rho]$ has been used in previous works to improve the {\em lower} bound for $c(2)$ (which plays a role in XC approximations such as SCAN\cite{SunRuzPer-PRL-15}) and for $c(N\le 60)$. Since\cite{SeiVucGor-MP-16,SeiBenKooGor-BOOK-22}
\begin{align}
    c(N)=\sup_{\rho\mapsto N}\Lambda_C[\rho], \qquad C=\lim_{N\to\infty} c(N)=\sup_{\rho}\Lambda_C[\rho],
\end{align}
improving the lower bounds for $c(N)$ amounts to find densities that give particular high values for $\Lambda_C[\rho]$. 

In Refs.~\citenum{SeiVucGor-MP-16,SeiBenKooGor-BOOK-22} it was observed that certain density profiles, such as a spherically-symmetric exponential, $\bar{\rho}(\rv)\propto e^{-r}$, have very high values of $\Lambda_C[\rho]$ already for small $N$, while other profiles, such as a sphere of uniform density, yield much lower values. In the next Sec.~\ref{sec:profiles}, we use our results of Sec.~\ref{sec:gradexp} to rationalise this observation.

\subsection{Why are some density profiles more challenging for the LO bound?}
\label{sec:profiles}
In Refs.~\citenum{SeiVucGor-MP-16} and~\citenum{SeiBenKooGor-BOOK-22}, values for $\Lambda_C[\bar{\rho}_{N,0}]$ were obtained by using different spherically-symmetric profiles $\bar{\rho}(r)$, with particle-number scaled densities $\bar{\rho}_{N,p}$ defined in Eq.~\eqref{rhoPNsca}. Notice that, due to Eq.~\eqref{eq:pdep},
$\Lambda_C[\bar{\rho}_{N,p}]$ is independent of $p$.
By inserting Eq.~\eqref{WexplicitN} into the definition of $\Lambda_C[\rho]$ of Eq.~\eqref{eq:LambdaLO}, we see that, for large $N$,
\bmath
\Lambda_C[\bar{\rho}_{N,p}]= -A_\infty\,-\,B_\infty\,\frac{I_2[\bar{\rho}]}{I_0[\bar{\rho}]}\,N^{-2/3}\,+\,...
\label{eq:LambdaLargeN}
\emath
Since $B_\infty >0$, the value $-A_\infty>0$ is approached from below as $N$ grows, indicating that the bcc Wigner crystal value is a local maximum for $\Lambda_C[\rho]$. Moreover, we see that density profiles with small values of the ratio $I_2[\bar{\rho}]/I_0[\bar{\rho}]$ will approach this maximum faster than density profiles for which this ratio is high.

\begin{figure}[ht]
\centering
\includegraphics[width=\columnwidth]{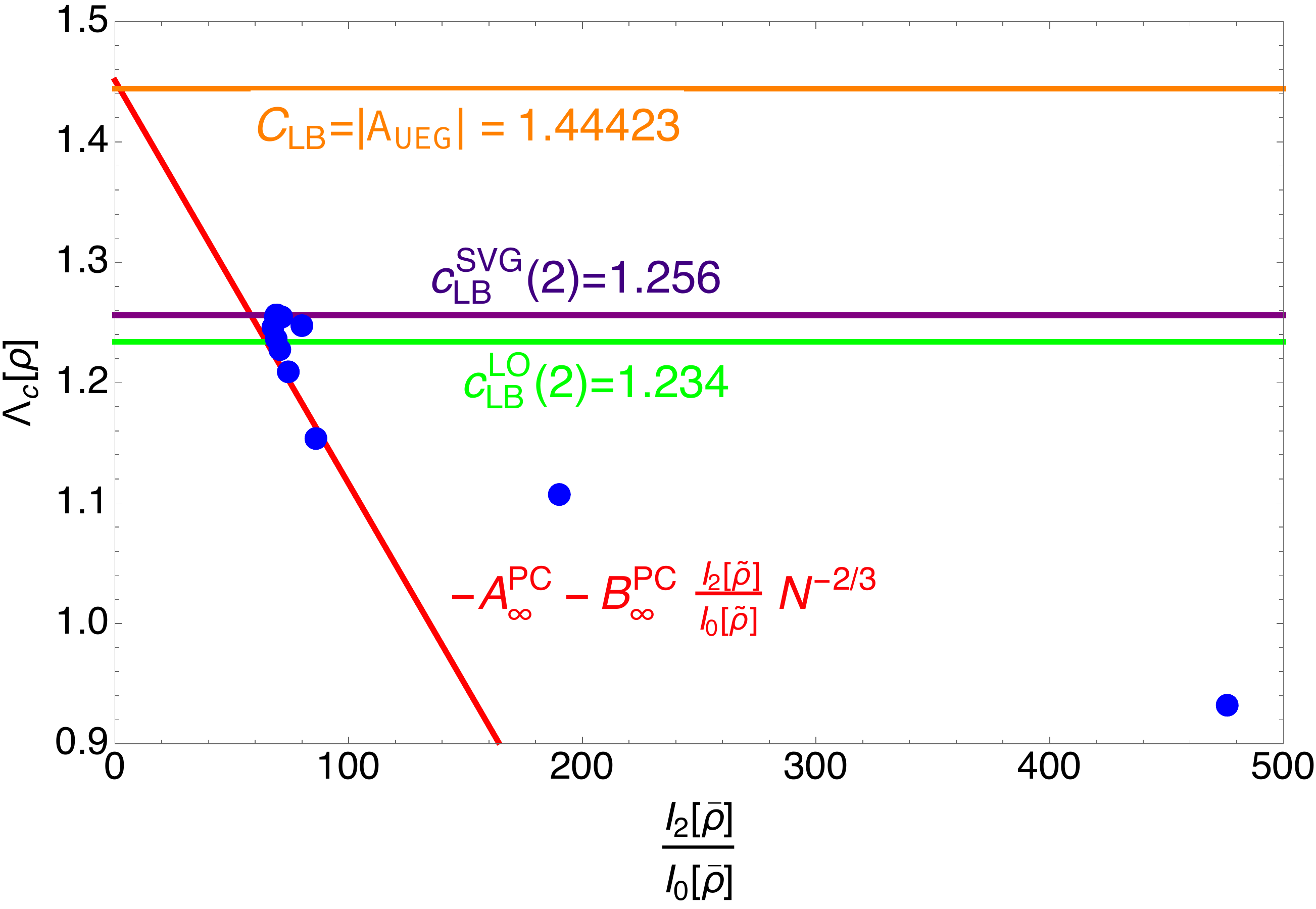}
\caption{$\Lambda_C[\bar{\rho}_{2,0}]$ of the different profiles from Table 1 of Ref.~\onlinecite{SeiVucGor-MP-16} plotted against the ratio $I_2[\bar{\rho}]/I_0[\bar{\rho}]$ of the GEA and LDA integrals, defined in Eqs.~\eqref{eq:I0}-\eqref{eq:I2}. The red straight line is the prediction from the PC model,\cite{SeiPerKur-PRA-00} see Eq.~\eqref{eq:LamPC}. {\color{black}The horizontal green and purple lines are the Lieb-Oxford lower bounds for $N=2$ obtained from respectively Ref.~\citenum{LieOxf-IJQC-81} and Ref.~\citenum{SeiVucGor-MP-16}, whereas the orange line is the lower bound for the general LO bound obtained from Ref.~\citenum{CotPet-arxiv-17} and Ref.~\citenum{LewLieSei-LMP-22}.}
Densities with an infinite $I_2$ integral have been excluded.}
\label{fig:profiles}
\end{figure}

Although the expansion of Eq.~\eqref{eq:LambdaLargeN} is valid for large $N$, the ratio $I_2[\bar{\rho}]/I_0[\bar{\rho}]$ is an excellent predictor for detecting profiles with high values of $\Lambda_C$, already for $N=2$. This is illustrated in Fig.~\ref{fig:profiles}, where the values of $\Lambda_C[\bar{\rho}_{2,0}]$ from Table 1 of Ref.~\citenum{SeiVucGor-MP-16}, are reported as a function of the corresponding ratio $I_2[\bar{\rho}]/I_0[\bar{\rho}]$. {\color{black}This ratio can thus provide good guidance in the choice of density profiles to improve the lower bound for the optimal constants $c(N)$.}

The simple PC model\cite{SeiPerKur-PRA-00} $W^{\rm PC}_{\infty}[\rho]=A^{\rm PC}_{\infty}\,I_0[\rho]+B^{\rm PC}_{\infty}\,I_2[\rho]$ yields a rough estimate for $\Lambda_C[\rho]$, see the straight line in Fig.~\ref{fig:profiles},
\bmath
\Lambda_C^{\rm PC}[\bar{\rho}_{N,p}] = -A^{\rm PC}_{\infty} - B^{\rm PC}_{\infty}\frac{I_2[\bar{\rho}]}{I_0[\bar{\rho}]}\,N^{-2/3}.
\label{eq:LamPC}\emath
The radial density profiles that fall on top of the PC model line in Fig.~\ref{fig:profiles} are $\rho(r)\propto (1+r)^{-n}$ with $n=4,5,6,7$ and 10, an empirical observation for which we do not have an explanation.

Some of the profiles $\bar{\rho}$ considered in Ref.~\citenum{SeiVucGor-MP-16} do not have a finite GEA integral $I_2$, and have been excluded from Fig.~\ref{fig:profiles}. One such profile is the ``droplet,'' corresponding to a sphere of uniform density. In this case, the integral $I_2$ diverges (see Appendix~\ref{app:droplet}), leading to a different behavior for large $N$ (liquid drop model\cite{SeiVucGor-MP-16}), namely 
\bmath
\Lambda_C[\bar{\rho}^{\rm Dro}_{N,p}] = -A_\infty+\,q_1\,N^{-1/3}\,+\,q_2\,N^{-2/3}\,+\,... 
\label{eq:LambdaLargeN-LDM}
\emath
where both $q_1$ and $q_2$ are negative.\cite{SeiVucGor-MP-16}

If instead of scaled density profiles we use the neutral atoms sequence, we have yet a different large-$N$ dependence, due to the asymptotic divergence of the $I_2$ integral discussed in Sec.~\ref{sub:na}, namely
\begin{equation}
    \Lambda_C[\bar{\rho}^{\rm na}_N]=-A_\infty-\frac{b_1^{\rm na}}{a_0^{\rm na}}\,N^{-2/3}\log(N)+...\; ,
\label{eq:LambdaLargeN-na}
\end{equation}
where $a_0^{\rm na}$ and $b_1^{\rm na}$ are positive constants appearing in Eqs.~\eqref{eq:I0na}-\eqref{eq:I2na}. 
Comparing with the asymptotic behavior of Eq.~\eqref{eq:LambdaLargeN}, Eq.~\eqref{eq:LambdaLargeN-LDM} seems to explain the empirical observation that $\Lambda_C$ values for uniform droplets approach the large-$N$ limit very slowly\cite{RasSeiGor-PRB-11,SeiVucGor-MP-16} and, Eq.~\eqref{eq:LambdaLargeN-na}, that also neutral atom densities are not particularly challenging for the LO bound.\cite{SeiBenKooGor-BOOK-22}

\subsection{The functional $E_{\rm el}[\rho]$}
\label{sec:TLB}
In this section, we consider the functional $E_{\rm el}[\rho]$
that appears in the strong-coupling limit of the adiabatic connection (AC) that has the M{\o}ller-Plesset (MP) perturbation series as expansion at weak coupling,\cite{SeiGiaVucFabGor-JCP-18,DaaGroVucMusKooSeiGieGor-JCP-20,DaaKooGroSeiGor-JCTC-22}
\begin{align}
E_{\rm el}[\rho] & =  \min_{\{\rv_1,...,\rv_N\}}\left\{\sum_{i<j =1}^{N}\frac1{|\rv_i-\rv_j|}-\sum_{i=1}^N\int\di^3r\frac{\rho(\rv)}{|\rv_i-\rv|}+U[\rho]\right\}, \nonumber \\
N & =  \int \rho(\rv) \di^3r.
\end{align}
The functional $E_{\rm el}[\rho]$ is the minimum electrostatic energy of a neutral system composed by $N$ identical point charges exposed to a classical continuous charge distribution with charge density $\rho(\rv)$ of opposite sign, and provides another lower bound\cite{SeiGiaVucFabGor-JCP-18,DaaGroVucMusKooSeiGieGor-JCP-20,DaaKooGroSeiGor-JCTC-22} to the SIL functional,
 $E_{\rm el}[\rho]\le W_\infty[\rho]$. Dividing again by $-I_0[\rho]$, in addition to Eq.~\eqref{eq:LambdaLO}, we also have
\bmath
\frac{-W_{\infty}[\rho]}{I_0[\rho]}\le\frac{-E_{\rm el}[\rho]}{I_0[\rho]}.
\label{eq:WinfBoundEel}\emath
The equality is reached for the case of the uniform electron gas (UEG) density,\cite{LewLieSei-PRB-19}
\bmath
\frac{-W_{\infty}[\rho_{\rm UEG}]}{I_0[\rho_{\rm UEG}]}\;=\;|A_{\infty}|\;=\;\frac{-E_{\rm el}[\rho_{\rm UEG}]}{I_0[\rho_{\rm UEG}]},
\emath
where $W_{\infty}[\rho_{\rm UEG}]$ is realised by a floating bcc Wigner crystal with uniform density,\cite{LewLieSei-PRB-19}
while $E_{\rm el}[\rho_{\rm UEG}]$ with any of the equivalent bcc Wigner crystal origins and orientations. The important point is that the two functionals have the same value.\cite{LewLieSei-PRB-19}

\begin{figure}[ht]
\centering
\includegraphics[width=\columnwidth]{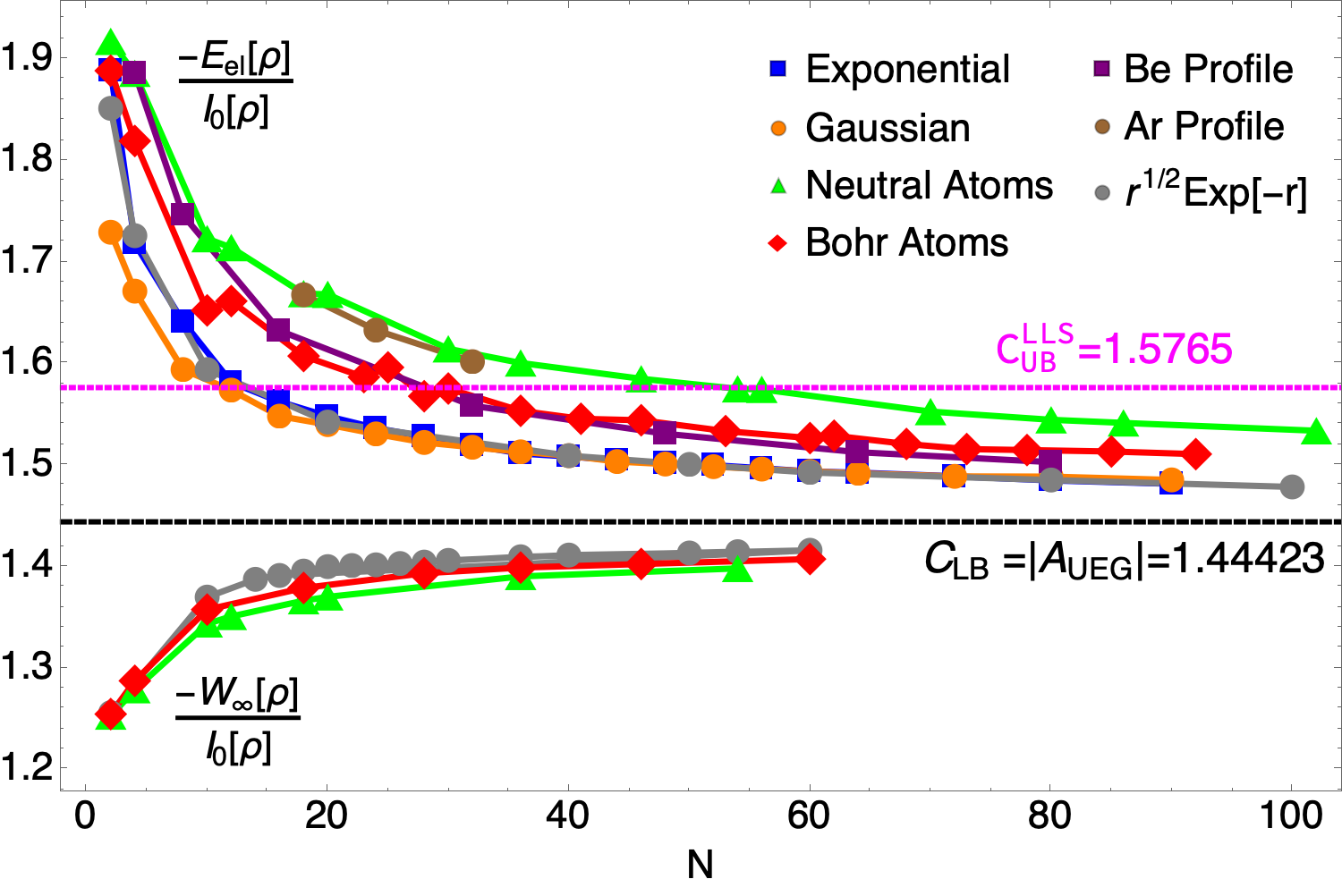}
\caption{ Numerical values for the functionals $-E_{\rm el}[\rho]/I_0[\rho]$ and $-W_\infty[\rho]/I_0[\rho]$. {\color{blue} Blue: Scaled exponential density.} 
{\color{orange} Orange: Scaled gaussian density.}
{\color{green} Green: Neutral atoms} 
{\color{red} Red: Bohr atoms} 
{\color{violet} Purple: Scaled Beryllium profile} 
{\color{brown} Brown: Scaled Argon profile} 
{\color{gray} Gray: $\rho^{\rm Sqc}_N(\rv)=\frac{2N}{15\,\pi^{3/2}}\,\sqrt{r}\,\ee^{-r}$.}\newline
Horizontal black line: $|A_{\rm UEG}|=|A_\infty|$ {\color{black} from Ref.~\citenum{LewLieSei-LMP-22} and Ref.~\citenum{CotPet-arxiv-17}.}  
{\color{magenta} Horizontal magenta line : the upper bound $C_{\rm UB}^{\rm LLS}$ proven in Ref.~\citenum{LewLieSei-LMP-22}.}}
\label{fig:sandwich}
\end{figure}

Values for $E_{\rm el}[\bar{\rho}_N^{\rm Sqc}]$ have been computed for neutral and Bohr atom densities, and for various particle-number scaled profiles in Ref.~\citenum{DaaKooGroSeiGor-JCTC-22}, and are combined, in Fig.~\ref{fig:sandwich}, with our present data to analyse the relationship with the LO bound. The figure suggests that  $-E_{\rm el}[\rho]/I_0[\rho]$ approaches its UEG value $|A_\infty|$ from above.

One could be tempted to think that this is a general feature. However, there is a simple counterexample with the property $-E_{\rm el}[\rho]/I_0[\rho]<|A_\infty|$: Consider the normalised density profile
\begin{equation}
\bar{\rho}(r)=\frac{(n+2)^{n+3}}{4 \pi \, \Gamma (n+3)}\,e^{-(n+2) r} r^n, \qquad n>0.
\label{eq:counter}
\end{equation}
As $n\to\infty$ this density approaches the Dirac measure of the unit sphere (a two-dimensional distribution, uniformly concentrated over the surface of the unit ball). For $N=1$, the value $E_{\rm el}$ remains finite while $I_0$ diverges, so their ratio will tend to 0. 
In such pathological cases then the LO bound becomes very loose, and $E_{\rm el}[\rho]$ in Eq.~\eqref{eq:WinfBoundEel} provides a much tighter lower bound to $W_\infty[\rho]$.

A caveat is that $E_{\rm el}[\rho]$ has also many local minima. For $W_\infty[\rho]$, this was not a problem, because even if one does not reach the global minimum, the computed value is still variational, providing a rigorous lower bound for $C$ in Eq.~\eqref{eq:LambdaLO}. For $E_{\rm el}[\rho]$, instead, a local minimum would provide an invalid lower bound to $W_\infty[\rho]$ in Eq.~\eqref{eq:WinfBoundEel}.
However, in our experience, the local minima of $E_{\rm el}[\rho]$ are all very close in energy, so in practice this might not be a severe problem.

\section{Computational details}
\label{sec:compdet}

\subsection{Densities}
\label{sec:denscalc}
For the neutral atoms, Hartree-Fock calculations were performed using \texttt{pyscf} 2.0.1.\cite{Sun-WIRCMS-18} An aug-cc-pVQZ\cite{KenDunHar-JCP-92} basis set was used, except for Ca (jorge-qzp\cite{JorCanCam-JCP-09}), Kr (cc-pVQZ\cite{KenDunHar-JCP-92}) and Xe (jorge-aqzp\cite{JorCanCam-JCP-09}). 

The densities of the Bohr atoms and $\sqrt{r}\,e^{-r}$ were computed analytically.

\subsection{Exchange functional $E_{\rm x}[\rho]$}\label{sec:Exfunc}
 For neutral atoms we used the Hartree-Fock exchange for the calculations described in Sec.~\ref{sec:denscalc} above. Although the Hartree-Fock exchange energy is not exactly the same as $E_{\rm x}[\rho]$ of KS DFT, the two values are very close, and for the qualitative study performed here the small differences should be unimportant. For example for $Z=10$ our $E_{\rm x}^{\rm HF}$ is equal to $-12.0847$, while the optimized effective potential (OEP) result, $E_{\rm x}^{\rm OEP}$, from Ref.~\citenum{ArgRedCanBur-PRL-22} is $-12.1050$. For $Z=36$ we have $E_{\rm x}^{\rm HF}=-93.805$ and $E_{\rm x}^{\rm OEP}=-93.833$.
 
 For the Bohr atoms, the data for $E_{\rm x}[\rho]$ are taken from Ref.~\citenum{ArgRedCanBur-PRL-22}.

 \subsection{Strong-coupling functional $W_\infty[\rho]$}
 Here we report the main details of the SCE calculations, with the full code to compute $W_\infty[\rho]$ for $N$ electrons in a given radial density profile available at \href{https://github.com/DerkKooi/jaxsce}{https://github.com/DerkKooi/jaxsce}.
 
All the densities considered here have spherical symmetry, and 
$W_\infty[\rho]$ was computed following the same procedure as in Refs.~\citenum{SeiGorSav-PRA-07,SeiVucGor-MP-16,VucLevGor-JCP-17} and~\citenum{SeiBenKooGor-BOOK-22}. 
This procedure relies on the radial optimal maps $f_i(r)$ of Ref.~\citenum{SeiGorSav-PRA-07}, which are known to provide either the exact $W_\infty[\rho]$ or a very close variational estimate of it.\cite{SeiDiMGerNenGieGor-arxiv-17} The calculation also requires a minimization on the relative angles (between electronic positions), which becomes very demanding as the number of electrons increases, due to the presence of many local minima. Overall, we can only be sure to provide a variational estimate of $W_\infty[\rho]$, which is obtained as
\begin{equation}
W_\infty[\rho]+U[\rho]=\int_{a_1}^{a_2}4\pi r^2 \rho(r)\, V_{\rm ee}(r)\, {\rm d}r,
\label{eq:WinftySCE}
\end{equation}
where $a_1$ and $a_2$ are defined below Eq.~\eqref{eq:cumulant}, and
\bmath
V_{\rm ee}(r) = \sum_{i < j} \frac{1}{|\rv_i(r) - \rv_j(r)|}. \label{eq:vee}
\emath
Here, in spherical polar coordinates, $\rv_i(r)=\big(f_i(r),\theta_i(r),\phi_i(r)\big)$ with $i=1,...,N$ is a set of $N$ strictly correlated position vectors, fixed by the distance $r$ of one of the electrons from the origin. The radial maps (or co-motion functions) $f_i(r)$,
\bmath
\underline{R}(r)=\big(f_1(r),f_2(r),f_3(r),\dots f_N(r) \big), \label{eq:maps}
\emath
with $f_1(r)=r$, are obtained from the density $\rho(r)$ via the cumulant function
\begin{equation}
\label{eq:cumulant}
N_e(r)=\int_0^r 4\pi x^2 \rho(x)\, dx
\end{equation}
and its inverse $N_e^{-1}(y)$, as detailed in Ref.~\citenum{SeiGorSav-PRA-07}. 
The integration limits in Eq.~\eqref{eq:WinftySCE} are $a_k=N_e^{-1}(k)$, with $k=1$ and 2 (but any pair of adjacent $a_k$ would work, due to cyclic properties of the maps\cite{SeiGorSav-PRA-07}).
For each value of $r\in[a_1,a_2]$, the set of relative angles $\{\theta_i(r),\phi_i(r)\}_{i=1,...,N}$ minimises the electron-electron interaction when the radial distances of the electrons from the nucleus are set equal to $\underline{R}(r)$ of Eq.~\eqref{eq:maps}.

The inverse cumulant $N_e^{-1}$ used in the calculation of the co-motion functions $\{f_i\}$ was either obtained analytically or by numerical inversion using the Newton method. The integration yielding $W_\infty[\rho]$ was performed on an equidistant grid between $a_1$ and $a_2$. For each $r\in[a_1,a_2]$, we find the minimizing relative angles $\{\theta_i(r),\phi_i(r)\}_{i=1,...,N}$  using the Broyden–Fletcher–Goldfarb–Shanno (BFGS) algorithm\cite{Bro-JAM-70,Fle-TCJ-70,Gol-MC-70,Sha-MC-70} in the \texttt{jaxopt.BFGS} function of \texttt{jaxopt}.\cite{jaxopt}. The number of grid points used for integration was 1025. Starting guesses of the optimal angles were obtained by minimization starting from 30000 sets of random angles at three different grid points: one close to the start of the interval, one in the middle of the interval and one close to the end of the interval. From these three starting points successively lower minima were obtained by sweeping forwards and backwards on the integration grid until convergence. For the last grid point, for which there is one less electron in the system, a separate angular minimization was performed from 30000 sets of random angles.

\section{Conclusions and Perspectives}
We have analysed gradient expansions of the weak- and strong-coupling functionals $E_{\rm x}[\rho]$ and $W_\infty[\rho]$ through the lens of particle-number scaling and neutral and Bohr atoms. Our main results are:
\begin{itemize}
    \item The compact representation in Eq.~\eqref{eq:BiN}, which allows to analyse an effective gradient expansion for all density sequences at the same time;
    \item The surprising symmetry in the effective gradient expansions of both functionals, which turn out to be very similar in magnitude but with opposite sign (see Fig.~\ref{fig:Bi});
    \item{A fresh look at the Lieb-Oxford bound for finite $N$, rationalising why some density profiles give better bounds than others (Sec.~\ref{sec:profiles}).}
\end{itemize}
Our findings can be used as constraints in building new XC functionals. For example, the fact that the coefficient of the gradient expansion should become positive at strong-coupling is a constraint ignored in all approximations. 

A question that seems to remain open is whether the larger (in magnitude) gradient expansion coefficient for exchange with respect to the one obtained by perturbing an infinite system with uniform density is due to the singular behavior of atomic densities with many electrons close to the nucleus, or whether this coefficient is simply different for finite systems. This question could be answered by computing $E_{\rm x}[\rho]$ for particle-number scaled densities, Eq.~\eqref{rhoPNsca}, starting from a given profile $\bar{\rho}$ with a finite $I_2[\bar{\rho}]$.  This calculation, however, requires a Kohn-Sham inversion for a given density, which is demanding for systems with many particles. For the functional $W_\infty[\rho]$, it seems that a scaled profile and neutral/Bohr atoms give very similar results, although we could only investigate here $N\le 60$. {\color{black} Another open question is the universality of $B_{x}$ and $B_{\infty}$ for finite densities, because we have only studied three atom-like density profiles. To provide more evidence for the density independence of $B_{x}$ and $B_{\infty}$, other density profiles, such as a scaled Gaussian density or diatomic molecules, should be studied in the future. The latter will also tell us about the transferability of our results from atoms to molecules.}

 \section*{Acknowledgments}
It is a pleasure to dedicate this work to John Perdew, who has been a wonderful mentor for some of us, and has pioneered all the problems and topics touched in this paper.

This work was funded by the Netherlands Organisation for Scientific Research under Vici grant 724.017.001. We thank Nathan Argaman, Antonio Cancio and Kieron Burke for the data of the exchange energy of Bohr atoms and for insightful discussions on the gradient expansion, and Stefan Vuckovic for discussions on the droplet data for the LO bound. We are especially grateful to Mathieu Lewin for suggesting to look at counterexamples of the kind of Eq.~\eqref{eq:counter}.
\section*{Data Availability}
The full code and all data are available at \href{https://github.com/DerkKooi/jaxsce}{https://github.com/DerkKooi/jaxsce} and in Ref. ~\citenum{jaxsce}.

\appendix

\section{Exchange for Bohr and neutral atoms} \label{app:Bohr}
In this appendix we analyze the results of Argaman et al.\cite{ArgRedCanBur-PRL-22} (hereafter denoted as ARCB) for the exchange energy of Bohr and neutral atoms in the light of our $\tilde{B}^{\rm Sqc}_{\rm x}(N)$.

Notice that ARCB define the Bohr atoms with external potential $-Z/r$ (with $Z=N$) rather than $-1/r$ as we did here. This corresponds to the scaling of Eq.~\eqref{eq:NIEions} for the densities,
\bmath\label{rhoBohrARCBsca}
\rho_N^{\rm BohrARCB}(r) &=& N^3\,\rho_N^{\rm Bohr}(Nr) \nonumber\\
&\approx& N^2\,\bar{\rho}^{\rm TFBohr}(N^{1/3}r)
\emath
where we have used Eq.~\eqref{scaBohr} in the second line. Due to Eq.~\eqref{rhoPNsca}, this is asymptotic particle-number scaling with $p=\frac13$, as in the neutral atoms case. Due to Eq.~\eqref{eq:scarel}, the first line of Eq.~\eqref{rhoBohrARCBsca} implies
\bmath
E_{\rm x}\big[\rho_N^{\rm BohrARCB}\big] = N\cdot E_{\rm x}\big[\rho_N^{\rm Bohr}\big]
\label{eq:BohrARCB}\emath
Our $\tilde{B}^{\rm Sqc}_{\rm x}(N)$ is insensitive to scaling, making visualization of the results independent of which definition is used.

ARCB use two different procedures for neutral and Bohr atoms to extract the final value of $B_{\rm x}$, and, on further inspection, rightly so (see below). For neutral atoms they only have values in a range of $N=Z$ similar to ours. They find the beyond-LDA exchange energies
\begin{equation}
\Delta E_{\rm x}^{\rm na}(N) = E_{\rm x}[\rho^{\rm na}_N]-A_{\rm x}\cdot I_0[\rho^{\rm na}_N]
\end{equation}
very accurately described by the simple two-parameter fit (Eq.~3 of ARCB)
\begin{equation}
\Delta E^{\rm na}_{\rm x}(N) = -0.0254 \, N \log(N) - 0.0560 N.
\end{equation}
For Bohr atoms, instead, for which ARCB have data for particle numbers up to $N=7590$, they fitted separately $E_{\rm x}[\rho^{\rm Bohr}_N]$ as
(Eq.~(11) of ARCB, where $A_{\rm x}\cdot a_0^{\rm Bohr}=-(\frac23)^{1/3}\frac4{\pi^2}=-0.354$).
\bmath \label{eq:ExBohrfit}
E_{\rm x}\big[\rho^{\rm Bohr}_{N}\big]&=&A_{\rm x}\cdot a_0^{\rm Bohr} N^{2/3}\\\nonumber
&+&e_1^{\rm Bohr}\,\log(N)+e_2^{\rm Bohr}\\\nonumber
&+&\frac{e_5^{\rm Bohr}\,\log(N)+e_6^{\rm Bohr}}{N^{2/3}}+\dots, \label{eq:ExBohr}
\emath
Here, our powers are smaller by a factor $N=Z$ due to Eq.~\eqref{eq:BohrARCB}.
Independently, they fitted $W^{\rm LDA}_0[\rho^{\rm Bohr}_N]=A_{\rm x}\cdot I_0[\rho^{\rm Bohr}_N]$ as 
\bmath \label{eq:LDABohrfit}
I_0\big[\rho^{\rm Bohr}_{N}\big] &=& a_0^{\rm Bohr} N^{2/3}\\\nonumber
&+& a_1^{\rm Bohr}\,\log(N)+a_2^{\rm Bohr}\\\nonumber
&+& \frac{a_3^{\rm Bohr}\,\log(N)+a_4^{\rm Bohr}}{N^{1/3}}\\\nonumber
&+& \frac{a_5^{\rm Bohr}\,\log(N)+a_6^{\rm Bohr}}{N^{2/3}}+\dots. \label{eq:LDABohr},
\emath\
{\color{black}until order $N^{-\frac{1}{3}}$ because $a_5^{\rm Bohr}$ and $a_6^{\rm Bohr}$ are very difficult to determine accurately from data. The values of the coefficients can be found in Ref.~\citenum{ArgRedCanBur-PRL-22}.} Taking the difference, we obtain the beyond-LDA exchange energy $\Delta E_{\rm x}^{\rm Bohr}(N)=E_{\rm x}\big[\rho^{\rm Bohr}_{N}\big]-A_{\rm x}\cdot I_0\big[\rho^{\rm Bohr}_{N}\big]$
\bmath
\nonumber\Delta E_{\rm x}^{\rm Bohr}(N) 
&=& \beta^{\rm Bohr}_1\,\log(N)+\beta^{\rm Bohr}_2 \nonumber\\
&+& \frac{\beta^{\rm Bohr}_3\,\log(N)+\beta^{\rm Bohr}_4}{N^{1/3}} \nonumber\\
&+& \frac{\beta^{\rm Bohr}_5\,\log(N)+\beta^{\rm Bohr}_6}{N^{2/3}}\,+\,...
\label{eq:BeyondBohr}\emath
(where $\beta_n=e_n-A_{\rm x}\cdot a_n$; note that $e_3=e_4=0$),
which for the first two terms gives,
\begin{equation}\label{eq:ExBohrBurke}
\Delta E_{\rm x}^{\rm Bohr}(N) = -0.03377 \, \log(N) - 0.05455,
\end{equation}
with the $\log(N)$ coefficient assumed in ARCB to be $\beta^{\rm Bohr}_1 \approx -\frac1{3\pi^2}=-0.03377$. (ARCB numerically find $e^{\rm Bohr}_1\approx-\frac7{27\pi^2}$ and $A_{\rm x}\cdot a^{\rm Bohr}_1\approx\frac2{27\pi^2}$.) (In their Eq.~9, ARCB define $B=-\beta_1$, not to be confused with our $B_{\rm x}$. Then, Eq.~17 in ARCB reads $\beta_1=\frac{27}{10}\beta_1^{\rm GEA}$.)

If, as an experiment, we fix (as ARCB did for neutral atoms) the coefficients by fitting $\Delta E_{\rm x}^{\rm Bohr}$ values for different limited ranges of $N$ ("small": $2\le N\le182$, "all": $2\le N\le7590$, "large": $1638\le N\le7590$), we find
\bmath
\Delta E^{\rm Bohr}_{\rm x}(N) =
\left\{\begin{array}{l}
-0.03256 \log(N) - 0.06550 \quad\text{(small $N$)} \\
-0.03298 \log(N) - 0.06420 \quad\text{(all $N$)} \\
-0.03318 \log(N) - 0.06277 \quad\text{(large $N$)}
\end{array}\right.
\label{TableIIrow-1}\emath
We see that the $\log(N)$ coefficient (with their extracted value being $\beta^{\rm Bohr}_1=-0.03377$) is fairly insensitive to the fitting range: Even with the small-$N$ range (similar to the one used by ARCB for neutral atoms) the error is only $3.6\%$.
Thus, it seems that extraction of the leading coefficient $\beta_1$ with the simple 2-parameter fit is relatively robust even when we have data available only in a smaller range of $N$. We should still stress, that, as suggested by the two different expansions of Eqs.~\eqref{eq:ExBohrfit} and \eqref{eq:LDABohrfit}, it is actually the exchange beyond its leading $N^{2/3}$ term that is really very accurately described by a two-parameter fit over a broad range of $N$. The LDA (Eq.~\eqref{eq:LDABohrfit}) has a much more complicated $N$ dependence, and its contribution beyond the leading $N^{2/3}$ term is not at all well described by a simple 2-parameter form. However, such contribution is also about one order of magnitude smaller with respect to the one of $E_{\rm x}$. Indeed, if we repeat the experiment of fitting the $N$-dependence of $E_{\rm x}[\rho_N^{\rm Bohr}]-A_{\rm x}\, a_0^{\rm Bohr} N^{2/3}$ on the $N\lesssim 150$ data, we get for the $\log(N)$ coefficient agreement with the full fit within 0.3\%.  So the 3.6\% error of $\beta_1$ in the first line of Eq.~\eqref{TableIIrow-1} is dominated by the LDA part. On the other hand, for neutral atoms, subtraction of the LDA diminishes the oscillations from the shell structure and makes data easier to fit.\cite{EllBur-CJC-09}

To extract $B_{\rm x}$ in $\Delta E_{\rm x}(N)=B_{\rm x}\cdot I_2(N)+...$ from values of $\Delta E_{\rm x}(N)$ given by Eq.~\eqref{eq:BeyondBohr}, 
we need the coefficient $b_1$ of the large-$N$ expansion
\bmath
I_2[\rho_N^{\rm Sqc}] = b^{\rm Sqc}_1\,\log(N) + b^{\rm Sqc}_2+...
\nonumber\emath
Its value has been determined analytically by ARCB, both for Bohr and neutral atoms. For the Bohr atoms they get $b^{\rm Bohr}_1=\frac{16}{(9\pi)^{1/3}}=5.25197$, which allows them to extract the value 
\bmath
B_{\rm x}^{\rm ARCB} = \frac{\beta^{\rm Bohr}_1}{b^{\rm Bohr}_1}
= \frac{-\frac1{3\pi^2}}{\;\frac{16}{(9\pi)^{1/3}}\;}
= -\frac{1}{16\pi (3\pi^2)^{1/3}} \approx -0.0064 \qquad 
\label{eq:BxARCB}\emath
listed in our Table~\ref{tab:values}. For neutral atoms, again comparing the coefficient of the $\log(N)$ term in the fit of $\Delta E_{\rm x}$ with the one of $I_2$, they obtain a very similar value, leading to the conjecture that $B_{\rm x}$ is the same for both series (as also apparent from our Fig.~\ref{fig:Bi}).

\begin{figure*}
\centering
\begin{subfigure}{.49\textwidth}
    \centering
    \includegraphics[width=.95\linewidth]{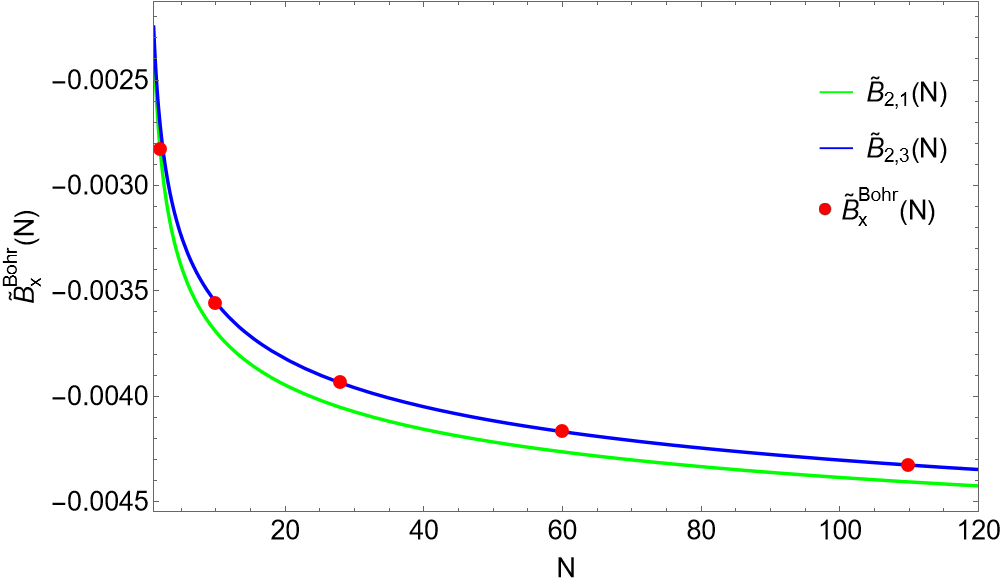}
\end{subfigure}
\begin{subfigure}{.49\textwidth}
    \centering
    \includegraphics[width=.95\linewidth]{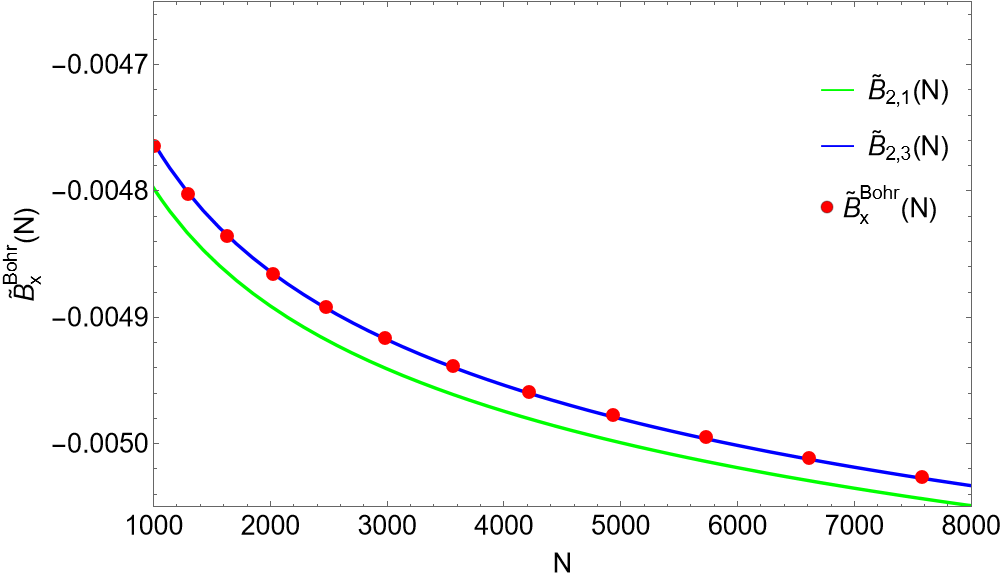}
\end{subfigure}
\caption{Exact data ({\color{red} red dots}) for $\tilde{B}_{\rm x}^{\rm Bohr}(N)$ and two different asymptotic expansion for it: $\tilde{B}_{2,1}(N)$ ({\color{green} solid green line}) and $\tilde{B}_{2,3}(N)$ ({\color{blue} solid blue line)} for small $N$ (left) and large $N$ (right). See Eqs.~\eqref{eq:BARCB21} and \eqref{eq:DKBSG23} for the full definition.}
\label{fig:Bfits}
\end{figure*}

The GEA integral $I_2[\rho_N^{\rm Bohr}]$, similarly to $I_0[\rho_N^{\rm Bohr}]$, is not well described by a simple two-parameter form, as repeating the fits in the different ranges of $N$ gives
\bmath
I_2[\rho^{\rm Bohr}_N] = \left\{\begin{array}{l}
4.7737 \log(N) + 28.1209 \qquad \text{(small $N$)} \\
4.8671 \log(N) + 27.7866 \qquad \text{(all $N$)} \\
5.0248 \log(N) + 26.5559 \qquad \text{(large $N$)}
\end{array}\right.
\nonumber\emath
where the $\log(N)$ coefficient (exact value\cite{ArgRedCanBur-PRL-22} $\frac{16}{(9\pi)^{1/3}}=5.2520$) has an error of around 10\% in the small-$N$ range and still 4\% for large $N$.
Although the next orders of $I_2[\rho^{\rm Sqc}_N]$ have not been studied in previous work, we will for now assume the expansion
\bmath
I_2[\rho^{\rm Sqc}_N] &=& b_1\,\log(N) + b_2 \nonumber\\
&+& \frac{b_3\,\log(N) + b_4}{N^{1/3}}
\emath
which matches the next order terms of $I_{0}$, see Eq.~\eqref{eq:LDABohrfit}.
Imposing the ARCB value $b_1=\frac{16}{(9\pi)^{1/3}}$ for the leading coefficient, we find by varying the other three coefficients $b_2$, $b_3$, and $b_4$ the accurate fit
\bmath\label{eq:GEAint4}
  \nonumber  I_2[\rho^{\rm Bohr}_N] &=& \frac{16}{(9\pi)^{1/3}}\,\log(N) + 23.7188\\
  && \,+\, \frac{1.2934\,\log(N)+4.1990}{N^{1/3}}.
\emath
Alternatively, by varying all coefficients $b_1$, $b_2$, $b_3$, and $b_4$, we obtain
\bmath
I_2[\rho^{\rm Bohr}_N] &=& 5.4072 \log(N) + 22.1207 \nonumber\\
&& + \frac{1.8551 \log(N)+5.6657}{N^{1/3}} \qquad \text{(small $N$)} \nonumber\\
I_2[\rho^{\rm Bohr}_N] &=& 5.3180 \log(N) + 22.9846 \nonumber\\
&& + \frac{1.5841 \log(N)+4.8467}{N^{1/3}} \qquad \text{(all $N$)} \nonumber\\
I_2[\rho^{\rm Bohr}_N] &=& 5.2302 \log(N) + 24.0881 \nonumber\\
&& + \frac{0.8543 \log(N)+4.9561}{N^{1/3}} \qquad \text{(large $N$)}
\nonumber\emath
An accurate estimate for the exact $b_1=\frac{16}{(9\pi)^{1/3}}=5.2520$ is recovered only when the fitting is limited to large $N$.

We can now use the fits of Eq.~\eqref{eq:ExBohrBurke}, \eqref{TableIIrow-1} and \eqref{eq:GEAint4} to find expressions for $\tilde{B}_{\rm x}(N)=\frac{\Delta E_{\rm x}}{I_2}$, via Eq.~\eqref{eq:BiN}. The two combinations that we use are
\bmath
\tilde{B}_{2,1}(N) \;=\;
\frac{-0.03377 \, \log(N) - 0.05455}{\frac{16}{(9\pi)^{1/3}}\,\log(N) + 23.7188},
\label{eq:BARCB21}\emath
obtained using {\color{black}Eq.~\eqref{eq:ExBohrBurke} plus the two leading orders of Eq.~\eqref{eq:GEAint4},}
and
\bmath
\tilde{B}_{2,3}(N) \;=\;
\frac{-0.03318 \log(N) - 0.06277}{\frac{16}{(9\pi)^{1/3}}\,\log(N) + 23.7188 \,+\, \frac{1.2934\,\log(N)+4.1990}{N^{1/3}}}, 
\label{eq:DKBSG23}:\emath
which combines the simple 2-parameter fit of $\Delta E_{\rm x}$ for large $N$ of Eq.~\eqref{TableIIrow-1} 
and the {\color{black} full} 3-parameter fit of $I_2$ of Eq.~\eqref{eq:GEAint4}. The number of parameters discussed here are those left free in the fit (we do not count the exact ones that were not varied in the fitting procedure). The subscript in the $\tilde{B}_{n,m}(N)$ are the numbers of the fitted (free) parameters in the numerator ($n$) and in the denominator ($m$).
We compare these two expressions against the data  for the exact $\tilde{B}_{\rm x}^{\rm Bohr}(N)$ in Fig.~\ref{fig:Bfits}. Notice that if we only use the two leading orders from $\Delta E_{\rm x}$ (Eq.~\eqref{eq:ExBohrBurke}) and from $I_2$ (Eq.~\eqref{eq:GEAint4}), we are still below the data even at $N$ as large as 7000. This shows that the asymptotic value is reached extremely slowly, {\color{black} although both fits capture the general trend well.}

\section{GEA integral of the droplet density}
\label{app:droplet}
In Ref.~\onlinecite{VucLevGor-JCP-17} the profile density for the droplets (spheres of radius 1 and uniform density) was approached as the limit $\alpha\to\infty$ of the radial profile
\begin{align}
    \bar{\rho}_\alpha(r)  =\frac{k(\alpha)}{1+e^{\alpha(r-1)}}, 
\end{align}
where the constant $k(\alpha)$ ensures that $\int_0^\infty 4 \pi r^2\bar{\rho}_\alpha(r)\,dr=1$. 

With the profile $\bar{\rho}_\alpha(r)$ it is possible to show explicitly the divergence of the integral $I_2$ for the droplet. In fact, lengthy but straightforward calculations lead to the result that, as $\alpha\to\infty$,  $I_2$ grows linearly with $\alpha$,
\begin{equation}
I_2[\bar{\rho}_{\alpha\to\infty}]=\left(\frac{4\pi}{3}\right)^{-2/3}\frac{18\pi}{5}\alpha+O(\alpha^0).
\end{equation}
For the droplet density profile, thus, the GEA expansion completely breaks down, explaining the different behavior\cite{RasSeiGor-PRB-11,SeiVucGor-MP-16} of Eq.~\eqref{eq:LambdaLargeN-LDM}.

\bibliographystyle{achemso}
\bibliography{bib_clean}

\end{document}